# An Experimental Information Gathering and Utilization Systems (IGUS) Robot to Demonstrate the 'Physics of Now'

Ronald P. Gruber[1]

*Stanford University Medical Center, Stanford, CA 94305, USA*

Ryan P. Smith

*Department of Physics, California State University - East Bay, Hayward, CA 94542, USA*

The past, present and future are not fundamental properties of Minkowski spacetime. It has been suggested that they are properties of a class of information gathering and utilizing systems (IGUSs). The past, present and future are psychologically created phenomena not actually properties of spacetime. A human is a model IGUS robot. We develop a way to establish that the past, present, and future do not follow from the laws of physics by constructing robots that process information differently and therefore experience different 'nows' ('presents'). We construct a customized virtual reality (VR) system which allows an observer to switch between present and past. This 'robot' (human with VR system) can experience immersion in the immediate past ad libitum. Being able to actually construct an IGUS that has the same 'present' at two different coordinates along the worldline lends support to the IGUS hypothesis.

## I. INTRODUCTION

In a prior AJP article, 'The physics of now,' J. Hartle remarked that there is neither a unique notion of space nor a unique notion of time.[1] specifically, the notion of past, present, and future are not properties of four-dimensional spacetime. Rather, these concepts are properties of a specific class of subsystems of the universe that process information and can usefully be called *information gathering and utilizing systems* (IGUSs). The concept of an IGUS has been utilized to better understand decoherent and consistent histories in quantum mechannics[2, 3] A human is one such robot. Their origin is to be found in how these IGUSs evolved or were constructed. Its 'present' is not a moment in spacetime. Rather, there is a 'present' at each instant along the robot's worldline.

Perhaps the easiest way of convincing oneself that the notions of past, present, and future do not follow from the laws of physics is to imagine constructing different robots that process information differently from one another. J. Hartle suggested that different notions of 'present' and even 'future' could be had for some of these robots with the aid of a virtual reality (VR) apparatus in which the data displayed was utilized differently than that of the human.

## II. A MODEL IGUS

We consider the human system as a model example of an IGUS. The system (Fig. 1) captures an image of its external environment at every proper time interval $t^*$. In this case, a stack of cards labeled *a, b, c, d, e, f*, etc. whose top member changes from time to time. The captured image is stored in register $P_0$ which constitutes the robot's present. Just before the next capture the image in $P_3$ is erased and images in $P_0$, $P_1$, and $P_2$ are shifted to the right making room for the new image in $P_0$. The registers $P_1$, $P_2$, and $P_3$ therefore constitute the robot's memory of the past. At each capture, the robot forgets the image in register P3. The robot uses the images in $P_0$, $P_1$, $P_2$ and $P_3$ in two processes of computation: C ~conscious, and U ~unconscious. The process U uses the data in all registers to update a simplified model or schema of the external environment. That is used by C together with the most recently acquired data in $P_0$ to make predictions about its environment to the future of the data in $P_0$, make decisions, and direct behavior. The robot may therefore be said to experience through C, the present in $P_0$, predict the future, and remember the past in $P_1$, $P_2$, and $P_3$.

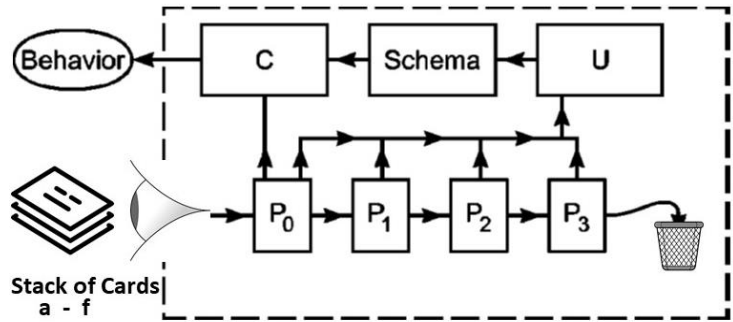

Fig. 1. At every proper time interval $t^*$, the robot captures an image of its external environment. In this case, this is of a stack of cards labeled *a, b, c, d, e, f,* etc. whose top member changes from time to time. The captured image is stored in register $P_0$ which constitutes the robot's present. Just before the next capture the image in $P_3$ is erased and images in $P_0$, $P_1$, and $P_2$ are shifted to the right making room for the new image in $P_0$. The registers $P_1$, $P_2$, and $P_3$ constitute the robot's memory of the past. Therefore, the robot may either experience the 'present' image in $P_0$ soon after arrival or later after it has passed to $P_3$. The information gathering and utilizing systems (IGUS) robot chooses how to best route and utilize its information. Figure modified from Ref. 1.

### III. ALTERNATIVE IGUSs

One of the IGUS robot examples is the *split screen* (*SS*) robot.[1] That robot has input to C computation from both the most recently acquired data in $P_0$ and from that in another register $P_j$, which was obtained a proper time $t_j = j \times t^*$ previously along its world line. Therefore, there is input to C computation from two distinct times. This disparity of information suggests that the robot's behavior will reflect the result of two sets of information that are temporally separated. The *SS* robot's present experience, it's 'now' (more accurately its 'present' moment) would consist of two times ($P_0$, $P_j$), equally vivid and immediate, i.e. two 'presents.'(one of which is technically a recent 'past'). It would remember the intermediate times ($P_1$, …, $P_{j-1}$), and the past ($P_{j+1}$, … $P_n$) through the U process of computation. J. Hartle suggested that the notion of what the 'present' would be for this *SS* robot could be experienced for a human by spending time in a virtual reality suit in which the data displayed was on an actual split screen. The result would be a *different perception of the physical 'now'* by being able to experience the 'present' and recent past simultaneously.

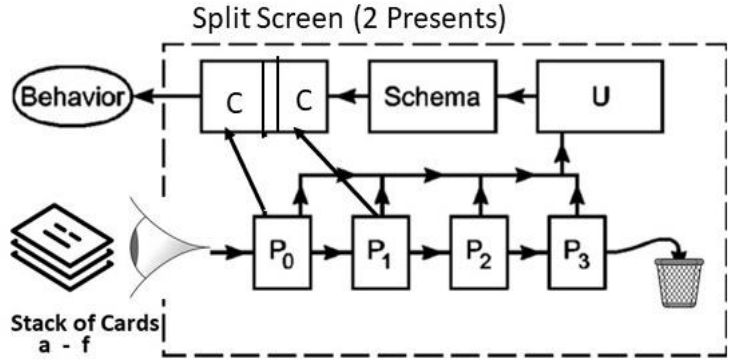

Fig. 2 Schematic of split screen IGUS robot. Memories from two worldline registers to two different Consciousness (C) computations appear on the split screen, giving the observer two 'presents' simultaneously. Figure is modified from Ref. 1.

Another example of an IGUS robot is the *always behind* (*AB*) Robot. This robot has input to C computation only from a particular register $P_k$, $K_0$ and the schema. That input is thus always a proper time $\tau_d = K\tau_*$ behind the most recently acquired data. It would remember the past stored in registers $P_{K+1}, \ldots P_n$. But also it would remember its future stored $P_0, \ldots P_{K-1}$ a time $\tau_d$ ahead of its present experience – reminiscent of premonitions of the future. Both the SS and AB robots have a different 'now' (present moment) from the model IGUS robot at each place along the worldline.

## IV. CONSTRUCTING AN IGUS 'ROBOT'

To construct the IGUS robot system, we developed a VR system consisting of a headset, mounted cameras, base stations and computer with the capability to handle the demands made on computation and memory that are required by caching large amounts of video data for a playback engine.[4] The camera used was a SP360 Video Action Camera from Kodak, run in the 180-degree configuration in a 1072x1072 resolution and mounted to the chest of the observer. The mounting of the cameras was an important aspect of creating immersion. If the cameras are in the headset the observer experiences his own head movement during replay and loses immersion. By having the camera on the upper chest a relatively stable visual background is provided for the observer's head to scan during replay (the 'past'). Subsequently, a stereoscopic variant of that camera was used in order to achieve greater immersion. The mobile headset camera was not used because a stable background scene is necessary when it comes time to replay. For robot mobility, all equipment fits into a backpack. Another important feature required was the use of base stations. Normally, these stations are stationary to provide tracking data of head position during replay. Since the robot ideally should be a mobile independent entity, an outrigger system to the backpack containing the computer was designed for the base station (Fig. 3a) where the cameras tracked head motion of the observer. This improvement greatly enhances the experience of immersion and distinguishes the experience from simply watching an 'instant replay' video. As an example, the observer is able to look around at different places in a 'past' scene period that she/he may not have looked during a 'present' scene period.

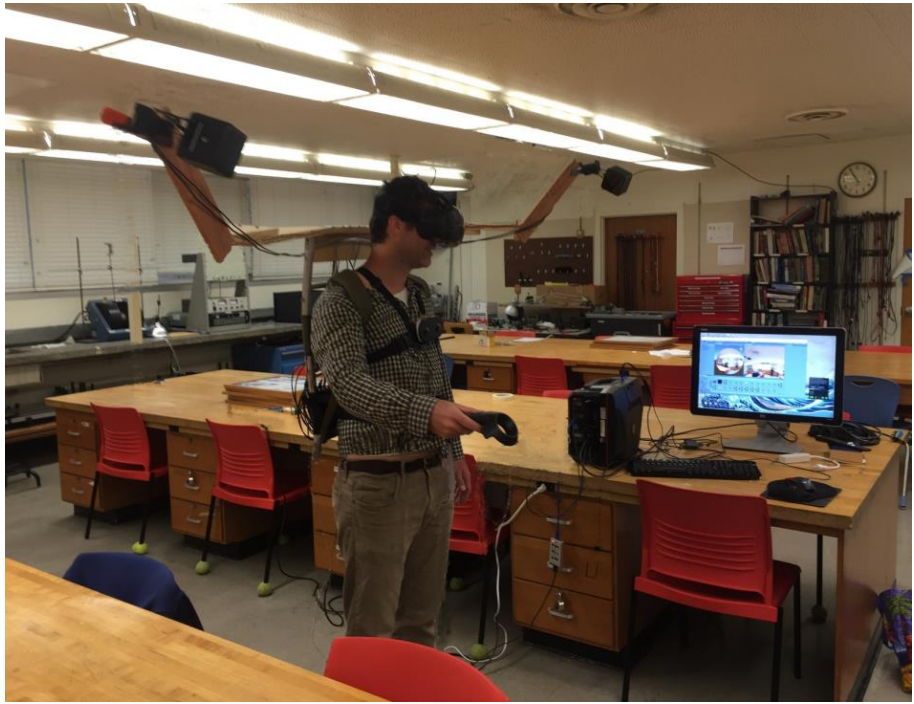

Fig. 3a. The human *intermittent behind (IB)* 'robot' has a VR system (target computer) worn as a backpack, base stations on retractable (telescopic) shoulder outriggers, and stereoscopic chest cameras for both viewing and recording. This design enhances the experience of immersion.

Initially we attempted construction of an *SS* virtual reality robot using a split headset screen. The interval between displays on the two hemi-screens (horizontally separated) was varied from 0.1 – 3.0 sec. However, upon construction, it was immediately apparent that the observer did not experience two simultaneous 'presents' because there was no sense of VR immersion when viewing a scene. In fact, it was confusing to the observer to see what are in effect, two slightly different images of the same object. In hindsight that is because VR requires complete immersion[5] which is impossible with a split screen. Achieving immersion demands a panoramic and, better yet, a 360 degree global scene displayed to the observer.

To create this immersion experience, we created a similar robot in which the observer is permitted to alternate between 'past' and 'present' panoramic screens ad libitum. By being able to switch between equally realistic time periods the observer experienced what was intended in the *SS* robot except in an alternating instead of a simultaneous manner. We should emphasize that it is not just replaying a video of the past events, but the user can "look around" at places where he perhaps did not look before, and the entire scene is made available for "experiencing," as distinct from watching an instant replay. Moreover, this robot version would be similar to the *AB* robot in that its 'present' would be of the immediate past but intermittently instead of continuously. Similar to the *SS* and *AB* robots this *intermittently behind* (*IB*) robot experiences a different 'present' along the worldline than the 'model IGUS,' but intermittently (not continuously) at the robot's choosing.

Like the *AB* robot, the *IB* robot experiences his/her future when in the 'past,' i.e. experiences awareness of future events ~~when~~ before leaving the 'past' and returning to the 'future.' Encouraged by recent work where Friedman et al. demonstrated that a VR suit enabled an observer to experience past events in a completely realistic manner in a fabricated (not a replay) 'past scene,' we were confident that this IB robot could be constructed.[4] In fact, the observer in that experiment was able to return to the scene of the 'past' on multiple occasions.

## V. TESTING THE 'ROBOT'

Seven participants were tested in this fashion, all of whom reported experiencing much the same phenomenon of 'presence' in past events – a key feature of immersion by VR.[5] A simple scene was utilized to allow careful analysis of the observer's behavior. It involved him/her watching an experimenter laying down three rows of dominoes (Fig 3b). The experimenter suggested to the observer that sometime during the third row when he wants to see the 'past' he/she should press a button to get a replay of the second row being laid out. When the observer pressed the button, the VR replay instantly initiated. The observer then experienced the extraordinarily realistic experience of 'being there' in the past. This is a similar experience described by the participants of the Friedman et al. experiment.[6] The participant perceived being at a different time with memory of where he/she was prior to replay. The participant was asked whether they agree or disagree with the two following specific statements presented to assess that perceptual phenomenon: 1) "seeing the second row of dominoes again was just as real as the first time;" and 2) "during VR replay of the second row of dominoes it seemed like I was 'there'." Participants unanimously agreed with both statements. This type of post-experiment questioning is more revealing than simply asking if there was a feeling of being in the 'past' because no participant has ever been in the past. The experience of 'being there' is the critical perceptual element that provides the observer with the sensation of past, present, and future as being .

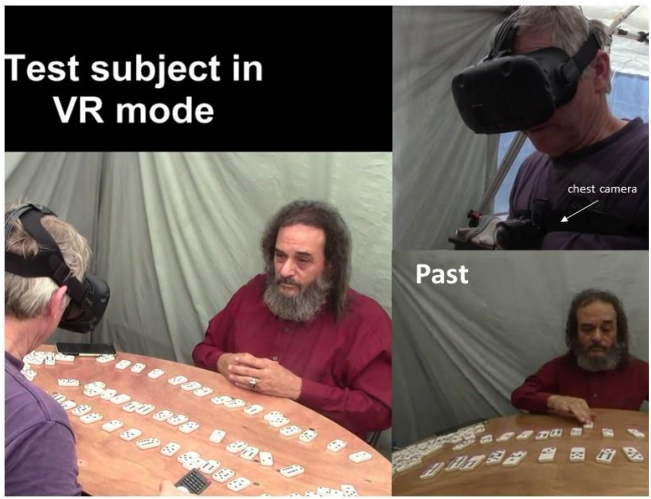

Fig. 3b  A simple scene was utilized to allow careful analysis of the observer's behavior.  It involved him/her watching an experimenter laying down three rows of dominoes The experimenter suggested to the observer that sometime during the third row when he wants to see the 'past' he/she should press a button to get a replay of the second row being laid out.  When the observer pressed the button, the VR replay instantly initiated.   The observer then experienced the extraordinarily realistic experience of 'being there' in the past. Note in lower right corner that the observer actually sees (realistically) the construction of the second row. Note the necessity of chest cameras in order to have a stable visual field upon replay.

In contrast with the model robot discussed in Section II, the *IB* robot *that chooses VR replay perceives being in the 'present' for the past scene as its coordinate along its worldline, and has a memory of its future in his/her actual location on the worldline.* Another way to describe this situation is that the *IB* robot perceives a particular 'present' twice, each at a different coordinate along the worldline. A worldline description of the 'present' for an *IB* robot is as follows in Fig. 4. It shows the worldline of an external object that is the source of its images such as the stack of cards in the prior figure. This source changes its shape at discrete instants of time delineated by ticks, passing through configurations *c, d, e, f, and g*. The configuration in each interval of the objects' worldline is labeled to the left.  The robot is permitted to utilize information as it chooses.  In this case the present from register e is experienced at two different points along the worldline, i.e., the IB robot experiences the same present twice.

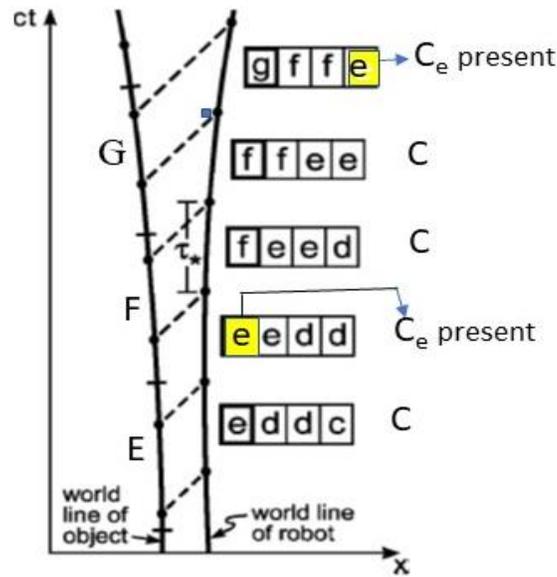

Fig. 4 A world-line description of the *intermittently behind (IB)* robot. Here the world line of an external object is the source of the images perceived by the robot, such as the stack of cards in the prior figure. This source changes its shape at discrete instants of time delineated by ticks, passing through configurations *c, d, e, f, and g*. The configuration in each interval of the objects' world line is labeled to the left. The robot is permitted to utilize information as it chooses. In this case the present from register *e* is experienced at two different points along the world line, i.e., the *IB* robot experiences the same present twice. Figure modified from Ref. 1.

In sum, the IGUS hypothesis suggests that the usual way in which a human thinks of the past present and future (as in the model IGUS) ~~are~~ is not the only conceivable way an IGUS can utilize temporal data in a four-dimensional physical world consistent with the known laws of physics. The *IB* robot is one such example that permits an observer to experience time in an unconventional way. Construction of that robot supports the IGUS hypothesis proposed by J. Hartle that a variant IGUS robot can be built which has a different way of experiencing the 'present.' Therefore, the 'present' is not unique[7] and not a property of Minkowski spacetime~~.~~ In other disciplines the uniqueness of the 'present' is questioned and it is considered to be a subjective, not an objective phenomenon.[8] Finally, it should be noted that this is not the first time VR has been used to help elucidate physical principles. Van Acoleyen and Van Doorsselaere recently used it to dramatize the effects of relativity using an educational approach similar to that of George Gamow's Mr. Tompkins.[9]

**ENDNOTES AND REFERENCES**